\documentclass[12pt]{article}
\usepackage{latexsym}

\def\tr{{\rm tr}}
\def\ket#1{\mid~\!\!\!{#1}~\!\!\rangle}
\def\bra#1{\langle~\!\!{#1}~\!\!\!\mid}

\def\IF{if and only if }
\def\QM{quantum mechanics }
\def\ch{characteristic }
\def\cH{{\cal H}}
\def\cR{{\cal R}}
\def\cS{{\cal S}}

\def\ON{orthonormal }
\def\li{linearly-independent }
\def\${\enskip$}
\def\#{\enskip$}
\def\cd{complete decomposition }

\begin{document}

{\bf \large \noindent On bipartite
pure-state entanglement structure\\
in terms of disentanglement}\\

\normalsize \rm

\begin{quote}
\noindent Fedor Herbut (E-mail:
fedorh@infosky.net and
fedorh@mi.sanu.ac.yu)\\

{\it \footnotesize
\noindent Serbian Academy of Sciences and
Arts, Knez Mihajlova 35,\\
11000 Belgrade, Serbia}\\

\normalsize

\rm \noindent Schr\"{o}dinger's
disentanglement [E. Schr\"{o}dinger,
Proc. Cambridge Phil. Soc. {\bf 31},
555 (1935)], i. e., remote state
decomposition, as a physical way to
study entanglement, is carried one step
further with respect to previous work
in investigating the qualitative side
of entanglement in any bipartite state
vector. Remote measurement (or,
equivalently, remote orthogonal state
decomposition) from previous work is
generalized to remote \li complete
state decomposition both in the
non-selective and the selective
versions. The results are displayed in
terms of commutative square diagrams,
which show the power and beauty of the
physical meaning of the (antiunitary)
correlation operator inherent in the
given bipartite state vector. This
operator, together with the subsystem
states (reduced density operators),
constitutes the so-called correlated
subsystem picture. It is the central
part of the antilinear representation
of a bipartite state vector, and it is
a kind of core of its entanglement
structure. The generalization of
previously elaborated disentanglement
expounded in this article is a
synthesis of the antilinear
representation of bipartite state
vectors, which is reviewed, and the
relevant results of Cassinelli et al.
[J. Math. Analys. and Appl., {\bf 210},
472 (1997)] in mathematical analysis,
which are summed up.
Linearly-independent bases (finite or
infinite) are shown to be almost as
useful in some quantum mechanical
studies as orthonormal ones. Finally,
it is shown that \li remote pure-state
preparation carries the highest
probability of occurrence. This singles
out \li remote influence from
all possible ones.\\

\end{quote}

\pagebreak

{\bf \large\noindent I. INTRODUCTION}

There are different {\it measures of
the amount} of entanglement in
bipartite states. In pure states they
all coincide. Hence, this is well
understood. But one may wonder the
measure of what is at issue; i. e.,
what is the structure of entanglement,
or what is its {\it qualitative side}.

According to Schr\"{o}dinger, the
natural way to investigate entanglement
is to perform {\it
disentanglement}:$^1$ It consists in
{\it measurements on the nearby
subsystem}. Since it is simultaneously
a measurement on the composite system,
the bipartite state becomes a mixed
one. As a consequence, one has an {\it
actual decomposition} (as opposed to a
potential or mathematical one) {\it of
the remote subsystem state}.

In previous work,$^2$ complete remote
measurement or, equivalently, complete
remote orthogonal state decomposition, was
studied as a first step in carrying out
Schr\"{o}dinger's program, and the concept
of twin observables was introduced. They
gave physical meaning to the so-called
correlated subsystem picture.$^3$

Mathematically, the optimal way to study
pure-state bipartite entanglement is to use
the antilinear operator representation of
the state vector. As it is well known, in
theoretical physics mathematics is
inextricably connected with physics. In the
mentioned previous work it turned out that
the antiunitary polar factor, the so-called
{\it correlation operator}, plays a central
role in establishing the concepts of twin
observables and remote measurement.
Naturally, this operator is endowed with
basic physical meaning.

The antilinear operator representation of
bipartite state vectors and the polar
factorizations of these operators are summed
up and shortly reviewed in Section 2.
Delving into the antilinear approach may
require some effort on part of some readers,
but it is pure-state bipartite entanglement
and not this author who made it optimal.
Eventually, the insight gained should make
it worth the effort.

In this article the physical content of the
correlated subsystem picture is extended one
step beyond remote measurement.

The organization of the rest of the article
goes as follows. In section 3 the relevant
purely mathematical results on
classification of all \li complete
decompositions  of any given density
operator (that with an infinite-dimensional
range included)$^4$ are shortly stated.
Besides, they are, to some extent,
elaborated in order to show that \li bases
can be almost as useful as \ON ones (to
encourage their use at least in entanglement
studies in \QM ). In section 4 the first
result of this paper, the generalized twin
observables, consisting of twin observables
and of extended twin observables, are
presented in the form of Theorem 1 and a
commutative square diagram. In section 5
selective (or specific-result)
nearby-subsystem measurement that gives rise
to so-called remote pure-state preparation
is payed special attention to in terms of
Theorem 2 and another square commutative
diagram. Besides, in Theorem 3 the physical
meaning of \li remote pure-state preparation
is clarified. In section 6 concluding
remarks point out the essential features of
the results.

As a technical remark, it should be noted
that by a basis (without further
specification) in a subspace is meant a
complete orthonormal set, i. e., one
spanning the subspace. We will also deal
with \li bases in linear manifolds (cf
Corollaries 1 and 3).\\

{\bf \large II. THE CORRELATED SUBSYSTEM
PICTURE}

The correlated subsystem picture is based on
the role of the antiunitary correlation
operator \$U_a\$ inherent in any bipartite
state vector \$\ket{\Phi}_{12}.\$ The
correlation operator is the antiunitary
polar factor of the antilinear
Hilbert-Schmidt operator \$A_a\$ that maps
the state space of subsystem \$1\$ into that
of subsystem \$2.\$ Such an operator, in
turn, gives an antilinear representation of
any given bipartite state vector
\$\ket{\Phi}_{12}$.

Antilinear operators were introduced in
physics from the mathematical literature$^5$
by Jauch.$^6$ They were utilized in Ref. 7
and in the first-step bipartite pure-state
studies.$^{2,8}$ The main result of these
was establishing the correlated subsystem
picture of \$\ket{\Phi}_{12}\$ (with the
twin observables) as a core of its
structure.\\

{\bf \large \noindent A. The mathematical
part}

Let \$\ket{\Phi}_{12}\# be a given state
vector of an arbitrary bipartite pure
state with a nearby (1) and a remote (2)
subsystem. Naturally, \$\ket{\Phi}_{12}\in
(\cH_1\otimes\cH_2),\$ where the tensor
factors are complex separable Hilbert
spaces.

The first notion that is being utilized is
that of the {\it partial scalar product}: If
\$\ket{\psi}_1\# is an arbitrary vector of
the nearby subsystem, then the partial
scalar product
$$\bra{\psi}_1\ket{\Phi}_{12}\in
\cH_2\eqno{(1)}$$ gives a vector in the
state space \$\cH_2\# of the remote
subsystem. It can be defined and evaluated
by introducing bases \$\{\ket{j}_1:\forall
j\}\subset \cH_1\# and
\$\{\ket{k}_2:\forall k\}\subset \cH_2\#
and expanding \$\ket{\Phi}_{12}\# in them:
$$\ket{\Phi}_{12}=\sum_j\sum_kf_{jk}\ket{j}_1
\ket{k}_2.\eqno{(2a)}$$ Then
\$\bra{\psi}_1\ket{\Phi}_{12}\# is
obtained in terms of the ordinary scalar
product in \$\cH_1$:
$$\bra{\psi}_1\ket{\Phi}_{12}=
\sum_j\sum_k\Big[f_{jk}\Big(\bra{\psi}_1
\ket{j}_1\Big)\Big]\ket{k}_2.\eqno{(2b)}$$
The point is, of course, that, as it is
straightforward to show, the rhs is always
defined (in case of infinite sums, one has
convergence), and the lhs is {\it
independent} of the choice of the subsystem
bases, and thus a well-defined
element of \$\cH_2$.\\

The next notion is that of the {\it
antilinear operator representation} of a
bipartite state vector \$\ket{\Phi}_{12}:\#
Relation (1) is actually an antilinear, i.
e., expansion-coefficients
complex-conjugating, map \$A_a\# of the
entire space \$\cH_1\# into \$\cH_2\#:
$$\Big(A_a\ket{\psi}_1\Big)_2\equiv
\bra{\psi}_1\ket{\Phi}_{12}\in
\cH_2.\eqno{(3)}$$ The operator \$A_a\#
defines its {\it adjoint} \$A_a^{\dag},\#
which maps in the antilinear way \$\cH_2\#
into \$\cH_1\#. This is done via the pair
of scalar products, that in \$\cH_1\# and
that in \$\cH_2$: $$\forall
\psi_1\in\cH_1,\enskip\forall\chi_2\in\cH_2:
\quad \Big(\chi_2,(A_a\psi_1)_2\Big)_2=
\Big((A_a^{\dag}\chi_2)_1,\psi_1\Big)_1
^*, \eqno{(4)}$$ where the asterisk
denotes complex conjugation. It is easy to
see that (4) defines adjoining as a linear
operation.

The operators \$A_a\# and \$A_a^{\dag}\#
are called Hilbert-Schmidt ones because
$$\tr (A_a^{\dag}A_a)_1<\infty,\quad \tr
(AA_a^{\dag})_2 <\infty.\eqno{(5a,b)}$$

The set of {\it all} antilinear
Hilbert-Schmidt operators mapping
\$\cH_1\$ into \$\cH_2\$ is a complex
separable Hilbert space, in which the {\it
scalar product} is defined as $$\forall
A_a,A_a':\quad \Big(A_a, A_ a'\Big)\equiv
\tr (A_a'^{\dag}A_a)_1.\eqno {(6)}$$ It is
straightforward to show that (3)
constitutes an {\it isomorphism} of the
complex separable Hilbert space of all
ordinary bipartite vectors
\$\ket{\Phi}_{12}\$ onto that of all
antilinear Hilbert-Schmidt operators
mapping \$\cH_1\$ into \$\cH_2$.

In the sense of this isomorphism, one can
speak of \$A_a\$ as {\it the antilinear
operator representative} of
\$\ket{\Phi}_{12}$.\\

It is known that the reduced density
operators \$\rho_1\equiv
\tr_2\Big(\ket{\Phi}_{12}
\bra{\Phi}_{12}\Big)\# and \$\rho_2\equiv
\tr_1\Big(\ket{\Phi}_{12}\bra{\Phi}_{12}
\Big)\# of any given bipartite state vector
\$\ket{\Phi}_{12},\# which describe the
respective subsystem states, have {\it equal
positive parts of their spectra}, i. e.,
their positive eigenvalues, together with
their multiplicities, coincide. Further, it
is known that if one expands
\$\ket{\Phi}_{12}\# in any eigen-sub-basis
\$\{\ket{r_i}_1:\forall i\}\# of \$\rho_1\#
spanning its range, then one obtains the
so-called {\it biorthogonal Schmidt
expansion}
$$\ket{\Phi}_{12}=\sum_ir_i^{1/2}\ket{r_i}_1
\ket{r_i}_2,\eqno{(7)}$$ where
\$\{r_i:\forall i\}\# are the positive
eigenvalues of \$\rho_1\# corresponding to
its mentioned eigenvectors, and
\$\{\ket{r_i}_2:\forall i\}\# turn out
necessarily to be eigenvectors of \$\rho_2\#
spanning its (equally dimensional) range.
Actually, one can write the spectral forms
as follows
$$\rho_1=\sum_ir_i
\ket{r_i}_1\bra{r_i}_1,\quad
\rho_2=\sum_ir_i
\ket{r_i}_2\bra{r_i}_2.\eqno{(8a,b)}$$

What the standard approach is lacking is
any expression of the {\it correlations}
between the two subsystems that the
bipartite state implies. This is where the
{\it antilinear} operator representation
of the bipartite state has a marked {\it
advantage}.\\

If one writes down the {\it polar
factorizations} of \$A_a,\# one obtains
$$A_a=U_a\rho_1^{1/2},\quad
A_a=\rho_2^{1/2}U_aQ_1\eqno{(9a,b)}$$ where
\$U_a\# is the antilinear unitary (or
antiunitary) {\it correlation operator},
which maps the (topologically) closed range
\$\bar \cR(\rho_1)\# of \$\rho_1\# onto
\$\bar \cR(\rho_2),\# that of \$\rho_2\#
(preserving the scalar product up to complex
conjugation). The Hermitian polar factors
are the positive-operator roots of the
corresponding reduced density operators, and
\$Q_1\# is the range-projector of
\$\rho_1.\# The operator \$U_a\# is uniquely
determined by \$A_a\# (i. e., by
\$\ket{\Phi}_{12}$), e. g., by
\$U_aQ_1=\tilde\rho_2^{-1/2}A_a\# (as
follows from (9b)), where \$\tilde \rho_2\$
is the reducee of \$\rho_2\$ in
\$\bar\cR(\rho_2).\$ (See also remark
beneath relation (12c).) (An elementary
discussion of polar factorization of linear
operators in one space is given in Ref. 9,
and a more general one in Appendix 4 of Ref.
2. The polar factorizations (9a,b) of
\$A_a\$ differ very little from this.)

It turns out that
$$\rho_2=\Big(U_a\rho_1U_a^{-1}\Big)_2Q_2
\eqno{(10)}$$ is valid, where \$Q_2\# is
the range-projector of \$\rho_2.\#
Utilizing \$U_a,\# the above Schmidt
expansion and the spectral forms can be
rewritten as follows:
$$\ket{\Phi}_{12}=\sum_ir_i^{1/2}
\ket{r_i}_1\Big(U_a\ket{r_i}_1\Big)_2,
\eqno{(11)}$$ and
$$\rho_1=\sum_ir_i\ket{r_i}_1\bra{r_i}_1,
\quad \rho_2=\sum_ir_i\Big(U_a\ket{r_i}_1
\Big)_2 \Big(\bra{r_i}_1U_a^{\dag}\Big)_2.
\eqno{(12a,b)}$$ (Note that
\$U_a^{\dag}=U_a^{-1}.$)  Actually,
$$\forall i:\quad
\ket{r_i}_2=U_a\ket{r_i}_1.\eqno{(12c)}$$
Thus, the correlation operator \$U_a\# can
be read off from the Schmidt biorthogonal
expansion (11) when the latter is explicitly
evaluated.

If \$\{\ket{j}_1;\forall j\}\# is a basis in
\$\cH_1,\# one can uniquely expand the
bipartite state, and, as easily seen, one
obtains $$\ket{\Phi}_{12}=
\sum_j\ket{j}_1\Big(A_a\ket{j}_1\Big)_2.
\eqno{(13a)}$$ The antilinear representation
\$A_a\$ of \$\ket{\Phi}_{12}\$ can be read
off from this because the antilinear
operator \$A_a\$ is continuous (cf Appendix
2 in Ref. 2); hence it is determined by its
action on a basis.

Relation (13a) can also be understood as
giving the inverse of isomorphism (3), i.
e., as determining the map \$A_a\enskip
\rightarrow\enskip\ket{\Phi}_{12}.\$ (It
is straightforward to show that the lhs of
(13a) does not depend on the choice of the
basis.)

If the basis in (13a) is an {\it eigenbasis}
of \$\rho_1,\# {\it then, and only then}, as
immediately seen from (9a), the general
expansion (13a) takes on the special form of
the biorthogonal Schmidt expansion (7).

The adjoint antilinear Hilbert-Schmidt
operators \$A_a^{\dag}\# also form a complex
separable Hilbert space in their turn with
the scalar product
$$\Big(A_a^{\dag},(A_a^{\dag})'\Big)\equiv
\tr (A_a'A_a^{\dag})_2.\eqno{(14)}$$ They
give the second antilinear operator
representation for bipartite vectors via the
isomorphism: $$\forall
\ket{\Phi}_{12}:\enskip \rightarrow \enskip
A_a^{\dag}:\quad \forall
\ket{\chi}_2:\enskip
\Big(A_a^{\dag}\ket{\chi}_2\Big)_1\equiv
\bra{\chi}_2\ket{\Phi}_{12}\in \cH_1.
\eqno{(15)}$$ Associating \$A_a^{\dag}\#
with \$A_a\# (cf (4)) is also an
isomorphism. (Any two of the mentioned three
isomorphisms of bipartite state spaces
multiply, i. e., give, when taken
one after the other, the third one.)\\

One has the following relations that are
symmetric to (9a)-(13a) in terms of the
adjoint antilinear operator representation
of \$\ket{\Phi}_{12}$:
$$A_a^{\dag}=U_a^{-1}\rho_2 ^{1/2},\quad
A_a^{\dag}=\rho_1^{1/2}U_a^{-1}Q_2.
\eqno{(16a,b)}$$

Further,
$$\rho_1=U_a^{-1}\rho_2U_aQ_1;\eqno{(17)}$$
$$\ket{\Phi}_{12}=\sum_ir_i^{1/2}\Big(U_a^{-1}
\ket{r_i}_2\Big)_1\ket{r_i}_2,\eqno{(18)}$$
where \$\{\ket{r_i}_2:\forall i\}\# is any
eigen-sub-basis of \$\rho_2\# spanning the
range of the latter, and \$\{r_i:\forall
i\}\$ are the corresponding (positive)
eigenvalues.

$$\rho_1= \sum_ir_i
\Big(U_a^{-1}\ket{r_i}_2\Big)_1\Big(
\bra{r_i}_2U_a \Big)_1=\sum_ir_i\Big(
U_a^{-1}(\ket{r_i}_2\bra{r_i}_2)U_a\Big)_1
,\eqno{(19a)}$$
$$\rho_2=\sum_ir_i\ket{r_i}_2\bra{r_i}_2.
\eqno{(19b)}$$ In general,
$$\ket{\Phi}_{12}=\sum_k
\Big(A_a^{\dag}\ket{k}_2\Big)_1\ket{k}_2,
\eqno{(20a)}$$ where \$\{\ket{k}_2:\forall
k\}\# is any basis in \$\cH_2.\# Again, it
is clear from (16a) that if this basis is an
eigenbasis of \$\rho_2,\# then and only
then, the general expansion (20a) takes the
special form of the biorthogonal Schmidt
expansion (18).

Finally, one has $$\rho_1=A_a^{\dag}A_a,
\quad\rho_2=A_aA_a^{\dag}.\eqno{(20b,c)}$$\\

The correlation operator \$U_a\#
establishes a striking {\it mathematical
symmetry} and close connection between the
two closed ranges \$\bar
\cR(\rho_s),\enskip s=1,2,\# for any
bipartite state vector
\$\ket{\Phi}_{12}.\# The pair of entities
\$\rho_1,U_a,\$ which is equivalent to
\$\ket{\Phi}_{12}\$ (cf (9a) and (3)), is
called {\it the correlated subsystem
picture} of the given bipartite state
vector. (Note that when one takes a state
vector \$\ket{\Phi}_{12}\$ instead of a
state \$\ket{\Phi}_{12}\bra{\Phi}_{12},\$
the former is informationally richer by
the choice of a fixed phase factor
\$e^{i\lambda},\enskip\lambda\in${\bf R},
which is arbitrary in the latter. This
choice is carried by \$A_a\$ or \$U_a.\$
Thus, \$U_a\$ and \$e^{i\lambda}U_a\$ with
the same \$\rho_1,\$ correspond to
\$\ket{\Phi}_{12}\$ and \$e^{i\lambda}
\ket{\Phi}_{12}\$ respectively.)

We have summed up in this Subsection the
mathematical part of the antilinear
representation of \$\ket{\Phi}_{12},\$ and
of the correlated subsystem picture. The
basic physical meaning of these was studied
in previous articles.$^{2,8}$ A summary is
given in the next subsection.\\

{\bf \large \noindent B. The physical part -
detectably-complete state-\\ compatible
observables}

Returning to the general expansion (13a),
it can be completed by
$$\rho_2=\sum_j\Big(A_a\ket{j}_1\Big)_2
\Big(\bra{j}_1A_a^{\dag}\Big)_2=\sum_jp_j
\ket{\phi_j}_2\bra{\phi_j}_2,
 \eqno{(13b)}$$ $$\forall j:\quad p_j\equiv
 ||\Big(A_a\ket{j}_1\Big)_2||^2,\eqno{(13c)}$$
 $$\forall j,\enskip p_j>0:\quad
 \ket{\phi_j}_2\equiv p_j^{-1/2}\Big(A_a
 \ket{j}_1\Big)_2.\eqno{(13d)}$$ Here
 \$p_j\$ is the probability that the event
 \$(\ket{j}_1\bra{j}_1\otimes 1)\$ occurs
 in nearby-subsystem measurement in
 \$\ket{\Phi}_{12},\$ and
 \$\ket{\phi_j}_2\$ is the state of the
 remote subsystem thus obtained, i. e., it
 is the result of so-called {\it remote
 preparation}. From the non-selective (or
 entire-ensemble) point of view, the physical meaning of (13a-d) consists
in the fact that these relations express a
{\it remote complete state decomposition} in
the antilinear representation. (In remote
incomplete state decomposition also mixed
states of the remote subsystem are obtained.
Such a remote decomposition is given rise to
by incomplete nearby-subsystem measurement,
i. e., by measurement of an observable with
degenerate eigenvalues.)

A detectably complete (see below) nearby
subsystem observable \$A_1\# that is
compatible with the nearby subsystem state,
i. e., that satisfies \$[A_1,\rho_1]=0,\$
shortly, a {\it state-compatible}
observable, has, on account of this
relation, as it is well known, a common
eigenbasis with \$\rho_1.\$ Let its
sub-basis spanning \$\bar\cR(\rho_1)\$ be
\$\{\ket{r_i}_1:\forall i\}\$ (cf (7) and
(8a)). Then the relevant partial spectral
form of \$A_1\$ is
$$A_1=\sum_ia_i\ket{r_i}_1\bra{r_i}_1
+Q_1^{\perp}A_1,\quad i\not=
i'\enskip\Rightarrow\enskip a_i\not=
a_{i'},\eqno{(21a)}$$ where
\$Q_1=\sum_i\ket{r_i}_1\bra{r_i}_1\$ is the
range projector of \$\rho_1,\$
\$Q_1^{\perp}\$ is the orthocomplemnentary
projector, and the sum in (21a) is the {\it
detectable part} (in \$\bar\cR(\rho_1)\$) of
\$A_1$.

By {\it detectably complete} is meant the
requirement in (21a), i. e., completeness
of the reducee \$\widetilde{A_1}=\sum_ia_i
\widetilde{\ket{r_i}\bra{r_i}}\$ of
\$A_1\$ in \$\bar\cR(\rho_1).\$ (For the
use of tilde cf \$\tilde\rho_2\$ in the
passage beneath (9a,b).)

When \$A_1\$ is measured (in an ideal way,
e. g.), it gives rise to the actual state
decomposition (empirically ensemble
decomposition): $$\rho_2=
\sum_ir_i\ket{r_i}_2
\bra{r_i}_2\eqno{(21b)}$$ (special case of
(13a-d)). Since the state vectors
\$\{\ket{r_i}_2:\forall i\}\# are
orthogonal (cf (7) and (8b)), (21b)
amounts to the same as if a detectably
complete remote-subsystem observable
(Hermitian operator) $$A_2=\sum_ia'_i
\ket{r_i}_2\bra{r_i}_2+Q_2^{\perp}A_2,\quad
i\not= i'\enskip\Rightarrow\enskip
a'_i\not= a'_{i'}\eqno{(21c)}$$ had been
measured in an ideal way. Here
\$Q_2=\sum_i\ket{r_i}_2\bra{r_i}_2\$ is
the range projector of \$\rho_2$.

The pairs of observables \$(A_1,A_2)\# are
called (physical) {\it twin observables},
the indirect measurement of \$A_2\# by
measuring \$A_1\# directly is called {\it
remote measurement}, and the twin
observables satisfy the symmetric
relations $$[A_1,\rho_1]=0,\quad
[A_2,\rho_2]=0;\eqno{(22a,b)}$$
$$A_2=\sum_ia'_i\Big(U_a(\ket{r_i}_1
\bra{r_i}_1)U_a^{-1}\Big)_2Q_2
+Q_2^{\perp}A_2, \eqno{(22c)}$$
$$A_1=\sum_ia_i\Big(U_a^{-1}(\ket{r_i}_2
\bra{r_i}_2)U_a\Big)_1Q_1+
Q_1^{\perp}A_1.\eqno{(22d)}$$

In (22c) it is assumed that \$A_1=\sum
a_i\ket{r_i}_1\bra{r_i}_1+Q_1^{\perp}A_1\$
is given, and \$A_2\$ is determined by it
(at least as far as the eigenvectors of the
detectable part of \$A_2\$ are concerned).
In (22d) the symmetrical assumption is made.
One should note that the undetectable parts
\$Q_1^{\perp}A_1\$ and \$Q_2^{\perp}A_2\$
are completely arbitrary (and so are the
distinct detectable eigenvalues of the twin
operator).

In Ref-s 2 and 8, it was assumed that {\it
the detectable spectra coincide}:
$$\forall i:\enskip a'_i=a_i.\eqno{(23a)}$$
Then
$$A_2=\Big(U_aA_1U_a^{-1}\Big)_2Q_2+
Q_2^{\perp}A_2,\eqno{(23b)}$$ and $$A_1=
\Big(U_a^{-1}A_2U_a\Big)_1Q_1+Q_1^{\perp}A_1
\eqno{(23c)}$$ are valid. In later
work,$^{10}$ twin observables with the
stronger requirement (23a) were called {\it
algebraic twin observables}. {\it
Relaxation} of the stronger requirement led
to the wider and more useful class of {\it
physical twin observables}.

 If one exchanges the roles
of subsystems \$1\$ and \$2,\$ one can
measure \$A_1\$ remotely by a direct
measurement of \$A_2$.

Thus, part of the {\it physical meaning}
of the {\it correlation operator} \$U_a\#
inherent in \$\ket{\Phi}_{12}\# is in the
following: When a detectably complete
nearby-subsystem observable \$A_1\# (cf
(21a)) that is compatible with the
nearby-subsystem state is measured in an
ideal way in selective measurement and
\$a_i\# is obtained as a result, then the
nearby subsystem is found in the state
\$\ket{r_i}_1,\# and the remote subsystem
is in the state
\$\ket{r_i}_2\equiv\Big(U_a\ket{r_i}_1
\Big)_2\# (conditional state). It is
obvious from (18) and (19a), that also the
symmetrical argument is valid. The
correlation operator (and its inverse)
give the corresponding {\it conditional
states} when selective ideal measurement
of state-compatible subsystem observables
is performed.

The remote measurement of a twin observable
\$A_2,\$ selective or non-selective, is one
and the same in every kind of measurement of
\$A_1:\$ in ideal measurement and in
second-kind (synonym: non-repeatable)
measurement (cf Subsection 6(B) in Ref.
2).\\

{\bf \large \noindent III.
LINEARLY-INDEPENDENT COMPLETE DECOMPOSITIONS
OF DENSITY OPERATORS}

{\it Definition 1:} A finite or countably
infinite set of vectors
\$\{\ket{\phi_i}:\forall i\}\# is said to be
{\it linearly independent} if $$\forall
i:\quad \ket{\phi_i}\notin \overline{span}
\{\ket{\phi_{i'}}:\forall i',i'\not=
i\},\eqno{(24)}$$ where by
\$"\overline{span}"\$ is meant the algebraic
and topological span, i. e., the set of all
linear combinations together with all their
limiting points. (It is a
subspace.)\\

One can define linear independence of a
finite sequence
\$\{\ket{\phi_i}:i=1,2,\dots ,d<\infty\}\#
of vectors by the weaker requirement:
$$\forall k,\enskip k\geq
2:\enskip\ket{\phi_k}\notin \mbox{span}\{
\ket{\phi_1},\dots ,\ket{\phi_{(k-1)}}\}.
\eqno{(25)}$$ Proof is given in Appendix A.
(Note that finite-dimensional linear
manifolds are subspaces, i. e., \$span=
\overline{span}$ in this case.)\\

{\it Definition 2:} If
\$\{\ket{\phi_i}:i=1,2,\dots
,d\leq\infty\}\# is a \li finite or infinite
sequence, \$\rho\# a density operator with a
\$d$-dimensional closed range, and if one
can write $$\rho =\sum_ip_i
\ket{\phi_i}\bra{\phi_i}, \eqno{(26a)}$$
where \$\forall i:\enskip p_i>0,\quad
\sum_ip_i=1,\# then one speaks of a {\it \li
\cd } of the density operator. (It is called
"irreducible decomposition" in Ref-s
11 and 4.)\\

We call "complete" those decompositions of
a density operator that cannot be
continued by further decomposing any term.
These are the pure-state decompositions
quantum mechanically. In a followup to
this article we turn to "incomplete"
decompositions, i. e., to mixed-or-pure
state decompositions quantum mechanically.

For (26a) the relation $$\bar\cR(\rho
)=\overline{span}
 \{\ket{\phi_i}:\forall i\}\eqno{(26b)}$$
is valid (cf Proposition 1 in
Ref. 11).\\

{\it Corollary 1:} Obviously, \ON sets are
special cases of linearly independent ones.
The latter possess some important properties
of the former. One of them is the following.
If \$\{\ket{\phi_i}:i=1,2,\dots ,d\leq
\infty\}\$ is a \li sequence, \$k\$ is an
integer not larger than \$d,\$ and
\$\{\ket{\phi_1},\dots ,\ket{\phi_k}\}\$ is
a subset of arbitrary elements in arbitrary
order, then it spans a \$k$-dimensional
subspace \$\cS_k,\$ and each vector in it
can be uniquely expanded in the set. This is
why the latter is called a {\it \li basis}
in \$\cS_k.\$
(See also Corollary 3 below.)\\

Proof is given in Appendix B.

Now we sum up those results on
density-operator decomposition from Ref. 4
the application of which forms the basis of
this work. They are further elaborated in
this section with a view to help
applications in quantum-mechanical studies.
(No heed is paid to the extent to which the
elaborations are possibly new with respect
to the mathematical
literature, cf e. g. Ref. 12.)\\

{\it Lemma:} {\it A)} Let \$\rho\$ be an
arbitrary given density operator, and let
\$d\$ be the dimension of its (finite or
infinite dimensional) closed range. Then
{\it all} \li sequences
\$\{\ket{\phi_i}:i=1,2,\dots ,d \}\$ that
determine a complete decomposition
$$\rho =\sum_ip_i\ket{\phi_i}\bra{\phi_i}
\eqno{(27)}$$ of \$\rho\# stand in a {\it
one-to-one relation} with the set of {\it
all} bases in \$\bar\cR(\rho )\$ each vector
of which is within \$\cR(\rho^{1/2})$:
$$\{\ket{e_i}:i=1,2,\dots
,d\leq\infty\}\subset \cR(\rho^{1/2}),
\eqno{(28)}$$ where \$d\$ is the dimension
of \$\bar\cR(\rho ).\$\\

{\it B)} The bijection from the set of all
bases (28) to all \li sequences that give
decompositions (27) - we call it the
Cassinelli-Vito-Levrero (CVL) bijection -
reads as follows:
$$p_i=\bra{e_i}\rho\ket{e_i}=
||\rho^{1/2}\ket{e_i}||^2>0,\quad
\ket{\phi_i}=
p_i^{-1/2}\rho^{1/2}\ket{e_i},\quad
i=1,2,\dots ,d\leq\infty .\eqno{(29a,b)}$$
The {\it inverse} CVL bijection is
$$\ket{e_i}=p_i^{1/2}\tilde\rho^{-1/2}
\ket{\phi_i},\quad i=1,2,\dots
,d\leq\infty ,\eqno{(30)}$$ where the
tilde denotes the reducee in
\$\bar\cR(\rho )$.\\

{\it C)} Finally, a state vector
\$\ket{\phi}\# can appear in a \li \cd of
\$\rho\# \IF $$\ket{\phi}\in\cR(\rho ),\quad
i=1,2,\dots ,d\leq\infty .
\eqno{(31)}$$\\

For proof see Theorem 1, Proposition 1,
and Remark 8 in Ref. 4.\\

In connection with the Lemma, one should
keep in mind the well-known (and easily
proved) relations $$\cR(\rho )\subseteq
\cR(\rho^{1/2})\subseteq
\bar\cR(\rho^{1/2})=\bar \cR(\rho
).\eqno{(32)}$$ In case of
finite-dimensional range, one has equality
all over. Contrarily, in case of
infinite-dimensional range, both subset
relations in (32) are {\it proper}.\\

{\it Corollary 2:} The CVL bijection is {\it
non-trivial} if and only if the basis (28)
is {\it not} an eigen-sub-basis of \$\rho$
(otherwise, it is the identity
map).\\

{\it Corollary 3:} Another property of
linearly-independent sequences parallelling
that of \ON ones is the following. If (27)
is a \li \cd of a density operator, then
each element \$\ket{\chi}\# from the range
\$\cR(\rho^{1/2})\# can be {\it uniquely
expanded} in the sequence
\$\{\ket{\phi_i}:i=1,2,\dots
,d\leq\infty\}$:
$$\ket{\chi}=\sum_i\alpha_i\ket{\phi_i}
\eqno{(33a)}$$ (cf (32)). Further, utilizing
the scalar product, one has the following
{\it compact} formula for the expansion
coefficients:
$$\alpha_i=p_i\Big[\Big(\bra{\phi_i}\tilde
\rho^{-1}\Big)\ket{\chi}\Big],\quad
i=1,2,\dots ,d\leq\infty\eqno{(33b)}$$ (cf
Lemma C)). In this sense, the sequence at
issue is a {\it \li basis} in
\$\cR(\rho^{1/2})$.

Note that the uniqueness of expansion
(33a) allows an arbitrary (hence, if
desired, a suitable) choice of the
probability distribution
\$\{p_i:i=1,2,\dots ,d\leq \infty
;p_i>0;\sum_{i=1}^dp_i=1\},\$ and the
definition of \$\rho\$ via (27). (But care
must be taken that \$\cR(\rho^{1/2})\$
contain \$\ket{\chi}.\$) Note, further,
that all \$d\$ probabilities \$p_i\$ must
be positive. Otherwise, \$\ket{\chi}\$
would not be expanded in the \li basis
\$\{\ket{\phi_i}:i=1,2,\dots
,d\leq\infty\}$.

Corollary 3 is proved in Appendix C.\\

{\it Corollary 4:} If (27) is a
linearly-independent complete decomposition
of a given density operator, then the weight
\$p_i\$ can also be expressed in the
following two ways:
$$p_i=\Big(\bra{\phi_i}\tilde\rho^{-1}
\ket{\phi_i} \Big)^{-1},\quad i=1,2,\dots
,d\leq\infty ,\eqno{(34)}$$ and
$$p_i=1\Big/\Big(\sum_k
(|\bra{k}\ket{\phi_i}|^2r_k^{-1})\Big),\quad
i=1,2,\dots ,d\leq\infty ,\eqno{(35a)}$$
where $$\rho
=\sum_kr_k\ket{k}\bra{k},\quad \forall k:
\enskip r_k>0\eqno{(35b)}$$ is a complete
spectral decomposition of \$\rho\#.

Further, one has
$$inf\{r_k:|\bra{\phi_i}\ket{k}|^2>0\}\leq
p_i\leq
max\{r_k:|\bra{\phi_i}\ket{k}|^2>0\},\enskip
i=1,2,\dots,d\leq\infty,\eqno{(36)}$$
where the "infimum" can be raplaced by
"minimum" if the range of \$\rho\# is
finite dimensional.\\

{\it Proof}: Expression (34) is obtained by
taking the square norm of both sides of
(30). Expression (35a) follows from (34)
when \$\ket{\phi_i}\# is expanded in the
eigen-sub-basis \$\{\ket{k}:\forall k\}\# of
\$\rho\$ (and eigenbasis of \$\tilde\rho
.)\# Finally, inequalities (36) are an
immediate consequence of
(35a).\hfill $\Box$\\

{\it Remark 1:} When a density operator
\$\rho\$ is given and a state vector
satisfies \$\ket{\phi}\in\cR(\rho )\$ ( cf
Lemma C)), then, in whatever
linearly-independent complete decomposition
of the former the latter appears, it has a
{\it unique} weight \$p,\$ which depends
only on \$\rho\$ and
\$\ket{\phi}\$ (cf (34)).\\

{\it Definition 3:} We call the weight \$p\$
from Remark 1 the {\it characteristic
weight} of \$\ket{\phi}\$ in \$\rho .\$ If
\$\ket{\phi}\notin \cR(\rho),\$ then
\$p\equiv 0$.

Note that if \$\ket{\phi}\in\cR(\rho ),\$
then \$p>0\$ (cf (30a)). Note, further,
that Remark 1 and Corollary 4 are a
completion of Lemma C.\\

{\it Remark 2:} For a possible positive
value of the \ch weight \$p,\# there may be
more than one corresponding state vector
\$\ket{\phi}\# in a \li complete state
decomposition (27) as seen from (29a),
because more than one state vector
\$\ket{f}\# can give one and the same
expectation value of \$\rho,\$ and each can
be the first \$\ket{e_1}\equiv\ket{f}\$ in a
basis etc.
(cf the Lemma).\\

{\it Corollary 5:} The \ch weight \$p\# of a
given state vector \$\ket{\phi}\in \cR(\rho
)\# satisfies the inequality:
$$p\leq \bra{\phi} \rho
\ket{\phi}.\eqno{(37)}$$ One has
\$p=\bra{\phi} \rho \ket{\phi}\# \IF
\$\ket{\phi} \# is an eigenvector of
\$\rho,\# and then \$p\# equals the
corresponding eigenvalue of the density
operator.\\

{\it Proof}: The inequality (37) follows
from (27) when one puts
\$\ket{\phi_1}\equiv\ket{\phi}\# in (27),
and one obtains $$\rho
=p\ket{\phi}\bra{\phi}+\sum_{i=2}p_i
\ket{\phi_i}\bra{\phi_i},\eqno{(38a)}$$ one
applies \$\ket{\phi}\bra{\phi}\# to both
sides, and one takes the trace (keeping in
mind, of course, that \$\tr
(\ket{\phi}\bra{\phi}\rho )=\bra{\phi}\rho
\ket{\phi}$): $$\bra{\phi}\rho \ket{\phi}=p+
\sum_{i=2}p_i|\bra{\phi}\ket{\phi_i}|^2 \geq
p.\eqno{(38b)}$$ One can see from (34) that
if \$\ket{\phi}\# is an eigenvector of
\$\rho\# corresponding to the eigenvalue
\$r,\# then \$p=r,\# and also
\$\bra{\phi}\rho \ket{\phi}=r=p.\#
Conversely, if \$p=\bra{\phi}\rho
\ket{\phi},\# then one can see from (38b)
that all vectors
\$\{\ket{\phi_i}:i=2,3,\dots \}\# must be
orthogonal to \$\ket{\phi}.\# Hence,
applying (38a) to \$\ket{\phi},\# it is seen
that the latter is an eigenvector of
\$\rho\# corresponding to the eigenvalue
\$r=p$. \hfill $\Box$\\

{\bf \large \noindent IV. REMOTE
LINEARLY-INDEPENDENT\\ COMPLETE STATE
DECOMPOSITION}

Let \$\ket{\Phi}_{12}\# be an arbitrary
bipartite state vector. Owing to the
Cassinelli et al. theory, summed up in the
Lemma, we can now easily sort out what
kind of local, i. e., subsystem
measurement gives rise to a \li \cd of the
opposite-subsystem state.\\

{\it Definition 4:} Since subsystem
measurement, by definition, excludes any
interaction between the measuring instrument
and the remote subsystem, we call any
influence of the former on the latter, which
is due exclusively to the quantum
correlations inherent in the bipartite
state, {\it remote} influence.\\

{\it Definition 5:} We call a nearby
subsystem observable \$A_1\# {\it relevant}
(for remote \li complete state
decomposition) if the following three
conditions are satisfied:

(i) $$[A_1,Q_1]=0,\eqno{(39)}$$ where
\$Q_1\# is the range projector of
\$\rho_1\equiv\tr_2\Big(
\ket{\Phi}_{12}\bra{\Phi}_{12}\Big)$. If
(39) is satisfied, then \$A_1\$ will be said
to be {\it range compatible}.

(ii) \$\tilde A_1,\# the reducee of
\$A_1\$ in \$\bar\cR(\rho_1)\$ if (39) is
satisfied, has a purely discrete and
non-degenerate spectrum.

(iii) The eigenbasis
\$\{\ket{e_i}_1:\forall i\}\$ of \$\tilde
A_1\$ in \$\bar\cR(\rho_1)\$ (which is
uniquely determined by \$\tilde A_1\$ up
to arbitrary phase factors and ordering)
is within \$\cR(\rho_1^{1/2}).\$ (This
requirement is always satisfied if the
dimension \$d\$ of \$\rho_1\$ is finite).

Further, we call a {\it basis}
\$\{\ket{e_i}_1:\forall i\}\$ in
\$\bar\cR(\rho_1)\$ that is entirely within
\$\cR(\rho_1^{1/2})\$ {\it relevant}.
Finally, we call a {\it class of observables
\$A_1\$ relevant} if it consists of relevant
observables that have {\it one and the same}
relevant basis \$\{\ket{e_i}_1:\forall i\}\$
(up to phase factors and ordering) as their
eigen-sub-basis in \$\bar\cR(\rho_1)$.\\

Evidently, the set of all relevant classes
of observables \$A_1\$ is in a simple
one-to-one relation with the set of all
relevant bases in \$\cR(\rho_1^{1/2})$.

If \$A_1\$ is {\it state-compatible}, i.
e., \$[A_1,\rho_1]=0,\$ then \$A_1\$
commutes also with every eigenprojector of
\$\rho_1,\$ and hence (39) is satisfied.
Namely, \$Q_1\$ is the sum of the
eigenprojectors corresponding to positive
eigenvalues. (If \$\cR(\rho_1)\$ is
infinite dimensional, we can assume that
\$A_1\$ is bounded, or, equivalently,
continuous, or equivalently, that its
spectrum is within a finite interval. We
can do this because the spectrum of
\$A_1\$ is arbitrary within the relevant
class of observables, i. e., it is
irrelevant for remote state
decomposition.)

In this case the reducee \$\tilde A_1\$ has
necessarily a purely discrete spectrum
(because it reduces in every eigen-subspace
of \$\rho_1,\$ and these are necessarily
finite dimensional due to the fact that the
corresponding eigenvalues add up to \$1).\$
Thus, in this case, requirement (i) is
necessarily fulfilled, and (ii) reads that
\$A_1\$ is a detectably complete observable,
i. e., that \$\tilde A_1\$ is complete.
Requirement (iii) is necessarily satisfied
because \$[A_1,\rho_1]=0\$  entails a common
eigenbasis of \$\tilde A_1\$ and \$\tilde
\rho_1.\$ Further, the corresponding
spectral form of \$\tilde \rho_1\$ is
simultaneously a complete decomposition of
it. Hence, according to the known result of
Hadjisavvas,$^{11}$ each of the eigenvectors
necessarily belongs to \$\cR(\rho_1^{1/2})$.
\\

{\bf Theorem 1:} {\bf A)} If a relevant
observable \$(A_1\otimes 1)\$ is measured in
the state \$\ket{\Phi}_{12},\$ it gives rise
to a remote \li \cd of the state \$\rho_2
\equiv\tr_1\Big(\ket{\Phi}_{12}
\bra{\Phi}_{12}\Big)\$:
$$\rho_2=\sum_ip_i\ket{\phi_i}_2
\bra{\phi_i}_2.\eqno{(40)}$$

Conversely, each mathematically possible
\li \cd of \$\rho_2\$ can be obtained in
this way.\\

{\bf B)} The mathematical way how \$A_1\$
determines (40) can be understood as a
bijection of the set of all classes of
detectably equivalent observables \$A_1,\$
or, equivalently, of all relevant bases
\$\{\ket{e_i}_1:\forall i\},\$ onto the set
of all linearly-independent complete
decompositions (40) (A$\searrow$D\enspace on
Diagram 1 below) that reads:
$$\forall i: \quad p_i=
\bra{\Phi}_{12}\Big(\ket{e_i}_1\bra{e_i}_1
\otimes
1\Big)\ket{\Phi}_{12}>0,\eqno{(41a)}$$
$$\forall i:\quad
\ket{\phi_i}_2=p_i^{-1/2}
\rho_2^{1/2}\Big(U_a\ket{e_i}_1\Big)_2,
\eqno{(41b)}$$ where \$\ket{e_i}_1\$ are
the eigenbasis vectors of \$\tilde
A_1=\sum_{i'}a_{i'}
\ket{e_{i'}}_1\bra{e_{i'}}_1.\$ Further,
\$U_a\$ is the antiunitary correlation
operator determined by \$\ket{\Phi}_{12}\$
(cf (9a,b) and the passage beneath it, as
well as the passage beneath (12c)).\\

{\bf C)} The inverse bijection
(A$\nwarrow$D on the diagram) has the
form: $$\forall i:\quad \ket{e_i}_1=
\Big(U_a^{-1}(p_i ^{1/2}\tilde
\rho_2^{-1/2} \ket{\phi_i}_2)\Big)_1.
\eqno{(42)}$$

All claims symmetric to those in A)-C) are
also valid:\\

{\bf D)} If \$(1\otimes A_2)=\sum_ja'_j
(1\otimes \ket{f_j}_2\bra{f_j}_2)+
(1\otimes Q_2^{\perp}A_2)\$ is the
relevant partial spectral form of an
arbitrary relevant observable \$A_2\$
(\$Q_2\$ being the range projector of
\$\rho_2\equiv\tr_1\Big(
\ket{\Phi}_{12}\bra{\Phi}_{12}\Big)\$),
its (non-selective) measurement causes a
remote linearly-independent complete state
decomposition $$\rho_1\equiv\tr_2\Big(
\ket{\Phi}_{12}\bra{\Phi}_{12}\Big)=
\sum_jq_j\ket{\chi_j}_1\bra{\chi_j}_1,\quad
\forall j:\enskip q_j>0,\enskip
\sum_jq_j=1.\eqno{(43)}$$ Each \li \cd of
\$\rho_1\$ can be obtained in this way.\\

{\bf E)} The bijection (C$\swarrow$B on the
diagram) taking the set of all relevant
classes of second-subsystem observables onto
that of all \li complete first-subsystem
state decompositions reads: $$\forall
j:\quad q_j\equiv
\bra{\Phi}_{12}\Big(1\otimes\ket{f_j}_2
\bra{f_j}_2\Big)\ket{\Phi}_{12}>0,
\eqno{(44a)}$$ $$\forall j:\quad
\ket{\chi_j}_1\equiv q_j^{-1/2}\rho_1^{1/2}
\Big(U_a^{-1}
\ket{f_j}_2\Big)_1. \eqno{(44b)}$$\\

{\bf F)} The inverse bijection
(C$\nearrow$B) is $$\forall j:\quad
\ket{f_j}_2\equiv\Big(U_a(q_j^{1/2}\tilde
\rho_1^{-1/2} \ket{\chi_j}_1)
\Big)_2.\eqno{(45)}$$
\\

{\bf G)} A bijection mapping all relevant
classes of observables \$(A_1\otimes 1)\$
onto that of all relevant classes of
observables \$(1 \otimes A_2)\$
(A$\longrightarrow$B on the diagram) is
$$\forall i:\quad\ket{f_i}_2\equiv
\Big(U_a\ket{e_i}_1\Big)_2.\eqno{(46a)}$$

The inverse bijection (A$\longleftarrow$B on
the diagram) is $$\forall
j:\quad\ket{e_j}_1\equiv
\Big(U_a^{-1}\ket{f_j}_2\Big)_1.
\eqno{(46b)}$$\\

{\bf H)} The product bijection
(C$\swarrow$B)$\circ$(A$\longrightarrow$B)
("$\circ$" meaning "after") is the
corresponding CVL bijection \$(A\downarrow
C);\$ and symmetrically, the product
bijection\enspace
(A$\searrow$D)$\circ$(A$\longleftarrow$B)
is the corresponding CVL bijection
\$(B\downarrow D)$.\\

{\bf I)} A bijection taking all \li \cd of
\$\rho_1\$ onto those of \$\rho_2\$
(C$\longrightarrow$D) is
$$\Big[U_a\Big(\rho_1=\sum_iq_i\ket{\chi_i}_1
\bra{\chi_i}_1\Big)U_a^{-1}\Big]_2Q_2,$$
giving, due to (10), $$\rho_2=\sum_jp_j
\ket{\phi_j}_2\bra{\phi_j}_2,$$ where
$$\forall j:\quad p_j\equiv
q_j,\eqno{(47a)}$$ $$\forall j:\quad
\ket{\phi_j}_2\equiv \Big(U_a\ket{\chi_j
}_1\Big)_2.\eqno{(47b)}$$

The inverse bijection (C$\longleftarrow$D on
the diagram) is symmetric to this (under the
exchange
of the two subsystems) mutatis mutandis.\\

{\bf J)} The square Diagram 1 summing up the
preceding items of Theorem 1 is {\it
commutative}, i. e., any two successive
bijections multiply into the corresponding
bijection on the Diagram.\\

\begin{center}
{\bf \large Commutative Square Diagram 1.\\A
Mathematical Framework for Remote\\
Linearly-Independent Complete State
Decompositions}
\end{center}
$${\bf\Huge A}\equiv\{\mbox{all relevant
classes of}\enskip A_1\}\qquad {\bf\Huge
B}\equiv\{\mbox{all relevant classes
of}\enskip A_2\}$$

$$\enskip {\bf \Huge
A}\quad\longrightarrow\quad
\quad\longleftarrow {\bf \Huge B}\enskip$$
$$\downarrow\searrow\star\enskip
\quad\qquad\star \swarrow\downarrow$$
\vspace{0.5cm}
$$\enskip\uparrow\nearrow\quad\qquad\qquad
\nwarrow\uparrow$$ $$\enskip {\bf \Huge
C}\quad\longrightarrow\quad
\quad\longleftarrow {\bf \Huge D}$$

$${\bf\Huge C}\equiv \{\mbox{all
linearly-independent complete
decompositions of}\enskip \rho_1\}$$
$${\bf\Huge D}\equiv\{\mbox{all
linearly-independent complete
decompositions of}\enskip \rho_2\}$$

\begin{quote}
\scriptsize \indent {\bf Caption.} Each
arrow goes from one of the sets (A,B,C,D)
towards another. It stands for the
corresponding bijection. Oppositely
oriented arrows denote mutually inverse
bijections. The diagram is {\it
commutative}, i. e., the successive
bijections combine into the displayed
corresponding one. For instance, taking
the bijection {\bf B}$\downarrow${\bf D}
after the bijection {\bf
A}$\rightarrow${\bf B} gives the bijection
{\bf A}$\searrow${\bf D}. The bijections
are given in detail in Theorem 1.

The imaginary vertical line cutting the
square into two equally wide halves makes
these completely symmetric (due to the
symmetry between the two subsystems in
\$\ket{\Phi}_{12}\$ ).

$\star$ The downward diagonal bijections
{\bf A}$\searrow${\bf D} and {\bf
C}$\swarrow${\bf B} have the {\it physical
meaning} of {\it remote \li state
decompositions}.
\end{quote}
\rm

{\it Proof} of Theorem 1.

{\it B)} To prove claim B, we take resort to
the relations (13a-d), which express the
general remote complete state decomposition.
Relation (13c) and (3) imply in our case
$$\forall i:\quad
0<p_i=||(A_a\ket{e_i}_1)_2||^2=||\bra{e_i}_1
\ket{\Phi}_{12}||^2=$$ $$
\Big((\bra{\Phi}_{12}
\ket{e_i}_1)_2,(\bra{e_i}_1\ket{\Phi}_{12}
)_2\Big)=\bra{\Phi}_{12}(\ket{e_i}_1
\bra{e_i}_1\otimes 1)\ket{\Phi}_{12}.$$
Further, (13d) and (9b) give \$\forall i:
\enskip\ket{\phi_i}_2=p_i^{-1/2}\rho_2^{1/2}
(U_a\ket{e_i}_1)_2$.

{\it C)} Claim C obviously follows from B in
view of the facts that both \$U_a\$ and
\$\tilde\rho_2^{1/2}\$ are non-singular on
\$\bar\cR(\rho_1)\$ and in
\$\bar\cR(\rho_2)\$ respectively.

{\it A)} The proof of claim A is a
consequence of part of the commutativity of
the square Diagram, viz., of the fact that
\$(A\searrow D)=(B\downarrow D)\circ
(A\longrightarrow B).\$ To see this, one
should keep in mind that
\$\rho_2=\Big(U_a\rho_1U_a^{-1}\Big)_2Q_2\$
(cf (10)) implies \$U_a\cR(\rho_1^{1/2})=
\cR(\rho_2^{1/2})\$ because the definition
\$\rho_1^{1/2}\rho_1^{1/2}=\rho_1\$ of the
square root, (10) and the well known
uniqueness of the square root lead to
\$\Big(U_a\rho_1^{1/2}U_a^{-1}\Big)_2Q_2
\Big(U_a\rho_1^{1/2}U_a^{-1}\Big)_2Q_2=
\rho_2,\$ and finally to
\$\Big(U_a\rho_1^{1/2}U_a^{-1}\Big)_2Q_2=
\rho_2^{1/2}.\$ Therefore,
\$\{\ket{e_i}_1:\forall i\}\subset\cR(
\rho_1^{1/2})\$ implies
\$\{(U_a\ket{e_i}_1)_2:\forall i\}\subset
\cR(\rho_2^{1/2})$.

Further, let us rewrite (41a) as $$\forall
i:\quad p_i=\bra{e_i}_1\rho_1\ket{e_i}_1=
\Big(\bra{e_i}_1U_a^{\dag}\Big)_2
\Big(U_a\rho_1U_a^{-1}\Big)_2
\Big(U_a\ket{e_i}_1\Big)_2=$$
$$\Big(\bra{e_i}_1U_a^{\dag}\Big) \rho_2
\Big(U_a\ket{e_i}_1\Big)_2.$$ (Complex
conjugation due to applying an antilinear
operator to the left is omitted because
the scalar product is a positive number.)

Comparing the last relation with (29a),
and (41b) with (29b), we see that the
claimed product of maps holds true. Since
the factors in the product are bijections,
so is the product itself (and its inverse
is the reverse product of the inverses).

Finally, on account of the fact that the
CVL bijection \$(B\downarrow D)\$ maps
onto the set of all \li complete state
decompositions (in \$\cR(\rho_2)),\$ the
same is valid for the remote state
decompositions \$(A\searrow D)\$ as
claimed.

The symmetric claims D, E, and F can be
proved symmetrically. Claim G is obviously
valid.

{\it H)} Claim H is an immediate consequence
of the products \$(C\swarrow B)=(A\downarrow
C)\circ (A\longleftarrow B),\$ which is the
symmetric relation of \$(A\searrow
D)=(B\downarrow D)\circ (A \longrightarrow
B)\$ (see proof of claim C).

Claim I is obviously valid. The final
claim J easily follows from the
multiplications proved for claim C.\hfill
$\Box$\\

{\it Remark 3:} In the special case when a
pair of range-compatible observables
\$A_s\enskip s=1,2,\$ are state-compatible,
then the CVL bijections \$(A\downarrow C)\$
and \$(B\downarrow D)\$ become
mathematically most simple, and are endowed
with physical meaning of actual orthogonal
decomposition of the states \$\rho_s,\enskip
s=1,2,\$ due to ideal measurement. If the
range-compatible observables are
state-incompatible, then
the CVL bijections are formal.\\

The {\it remote \li \cd } (40), caused by
the direct subsystem measurement of
\$(A_1\otimes 1),\$ {\it has the physical
meaning of actual decomposition} of
\$\rho_2.\$ This is so because, when the
measurement interaction is over, the
tripartite pure state vector has undergone
the change
$$\ket{0}_{MA}\ket{\Phi}_{12}=\ket{0}_{MA}
\Big[\sum_i\ket{e_i}_1[\rho_2^{1/2}(U_a
\ket{e_i}_1)_2]\Big]\rightarrow$$ $$
\sum_i\ket{i}_{MA}\ket{e'_i}_1
\Big[\rho_2^{1/2}\Big(U_a\ket{e_i}_1\Big)_2
\Big]=$$
$$\sum_ip_i\ket{i}_{MA}\ket{e'_i}_1
\ket{\phi_i}_2,\eqno{(48)}$$ where
\$\ket{0}_{MA}\$ is the initial state vector
of the {\it measuring apparatus}, and
\$\{\ket{i}_{MA}:\forall i\}\$ is the \ON
set of so-called "pointer positions": the
state vectors in it display the results
\$\{a_i:\forall i\}\$ in the measurement of
\$A_1.\$ The state vectors \$\ket{e_i'}_1\$
equal the initial state vectors
\$\ket{e_i}_1\$ if the measurement is a
non-demolition (repeatable) one, and they
differ otherwise. (Remember that we have
complete measurement.) Anyway, according to
(13a-d) (where now "$1$" is to be replaced
by "$(MA+1)$"), the final tripartite state
gives rise to the remote state decomposition
\$\rho_2=
\sum_ip_i\ket{\phi_i}_2\bra{\phi_i}_2\$.
After reading the results on the measuring
apparatus, i. e., after so-called {\it
objectivization},$^{13}$ this decomposition
becomes {\it actual} (in contrast to the
infinitely-many mathematically possible
so-called "potential" decompositions).

Returning to Theorem 1, and the square
Diagram 1, we can say that the latter
displays an extended physical meaning (with
respect to that in Subsection 2.2)
of the correlated subsystem picture.\\

{\it Definition 6:} Pairs of opposite
subsystem observables \$(A_1,A_2)\$
satisfying \$[A_s,Q_s]=0,\enskip s=1,2,\$
that are relevant (cf Definition 5) and can
be written either as \$A_1\$ and
\$A_2=\sum_ia'_i(U_a\ket{e_i}_1)_2(\bra
{e_i}_1U_a^{\dag})_2 +Q_2^{\perp}A_2\$ or as
\$A_1=\sum_ia_i(U_a^{-1}\ket{f_i}_2)_1
(\bra{f_i}_2U_a)_1+ Q_1^{\perp}A_1\$ and
\$A_2\$ (depending on the choice of the
nearby and the remote subsystems), which
amount to the same, can be called {\it
generalized twin observables}. If
\$[A_s,\rho_s]=0,\enskip s=1,2,\$ are {\it
not valid}, then one is dealing with {\it
extended twin observables}.\\

Diagram 1 displays the physical meaning of
the correlated subsystem picture that
includes the extended twin observables (in
addition to the special case of twin
observables).\\

{\it Corollary 6:} The relevant classes of
nearby-subsystem observables give an {\it
alternative classification} of all \li
complete decompositions of \$\rho_2.\$ This
can be extended to any density operator
\$\rho ,\$ if it becomes \$\rho_2\$ by
so-called purification, i. e., by
constructing a bipartite state vector
\$\ket{\Phi}_{12}\$ that implies the initial
density operator as its second-subsystem
reduced density
operator.\\

One should note that, from the point of view
of mathematical physics, this classification
has an {\it advantage} over that of
Cassinelli et al.$^4$ (cf the Lemma above)
consisting in the fact that the classifying
entities and the details of the connection
between them and the \li complete
decompositions has a clear physical meaning
in terms of the antilinear operator
representation of \$\ket{\Phi}_{12},\$ and
its polar
factorization (cf section 2).\\

{\it Remark 4:} In Ref. 14, the approach of
this section was indicated (without the
antilinear operator representation) for
finite-dimensional ranges. It was pointed
out that this can lead to generating \li \cd
of states even at space-like separation.\\

{\bf \large \noindent V. THE SELECTIVE
ASPECT OF REMOTE LINEARLY-INDEPENDENT
COMPLETE STATE DECOMPOSITION: REMOTE STATE
PREPARATION}

The selective or one-result sub-ensemble
aspect of complete subsystem measurement
that gives remote \li complete state
decomposition was only implicitly given so
far. Now we make it explicit. It is an
immediate consequence of Theorem 1.

{\bf Theorem 2:} Also the selective (or the
one-result sub-ensemble) aspect, i. e., the
{\it remote preparations} of pure states
that are part of a \li complete state
decomposition, can be displayed in a
commutative square diagram as below.  The
symbols on Diagram 2 have the following
meaning.

{\bf A} is the set of all state vectors
\$\ket{e_k}_1\$ from \$\cR(\rho_1^{1/2})\$
(equivalently, the set of all corresponding
atomic events or ray projectors
\$\ket{e_k}_1\bra{e_k}_1$). {\bf B} is the
set of all state vectors \$\ket{f_n}_2\$
from \$\cR(\rho_2 ^{1/2}).\$ {\bf C} is the
set of all state vectors \$\ket{\chi_n}_1\$
from \$\cR(\rho_1).\$ Finally, {\bf D} is
the set of all state vectors
\$\ket{\phi_k}_2\$ from \$\cR(\rho_2)$.

The bijection {\bf A}$\longrightarrow${\bf
B} comes about by application of the
antiunitary correlation operator \$U_a,\$
which is determined by the given bipartite
state vector \$\ket{\Phi}_{12}.\$ The
inverse bijection {\bf
B}$\longleftarrow${\bf A} is \$U_a^{-1}.\$
{\bf A}$\downarrow${\bf C} is
\$p_k^{-1/2}\rho_1^{1/2}.\$ The inverse is
{\bf C}$\uparrow${\bf
A}$=p_k^{1/2}\tilde\rho_1^{-1/2},\$ where
the tilde denotes the reducee to the range.
{\bf B}$\downarrow${\bf D} is
\$p_n^{-1/2}\rho_2^{1/2}.\$ The inverse is
{\bf D}$\uparrow${\bf
B}$=p_n^{1/2}\tilde\rho_2^{-1/2}.\$ {\bf
C}$\longrightarrow${\bf D} is \$U_a.\$ The
inverse bijection {\bf
C}$\longleftarrow${\bf D} is \$U_a^{-1}.\$

The diagonal arrows, which have the physical
meaning of remote state preparation, are the
following. {\bf A}$\searrow${\bf D} is
\$p_k^{-1/2}\rho_2^{1/2}U_a,\$ or,
equivalently, \$p_k^{-1/2}A_a.\$ The inverse
is {\bf A}$\nwarrow${\bf
D}$=p_k^{1/2}\tilde\rho_1^{-1/2} U_a^{-1}.\$

{\bf C}$\swarrow${\bf B} is
\$p_n^{-1/2}\rho_1^{1/2}U_a^{-1},\$ or,
equivalently, \$p_n^{-1/2}A_a^{\dag}.\$ The
inverse is {\bf C}$\nearrow${\bf
B}$=p_n^{1/2}\tilde\rho_2^{-1/2}U_a$.\\

\begin{center}
{\bf \large Commutative Square Diagram 2.\\
Remote Pure-State Preparation.\\ (The
Selective Aspect of\\
Remote Linearly-Independent Complete State
Decompositions)}
\end{center}
$${\bf\Huge A}\equiv\{\mbox{all state
vectors}\enskip
\ket{e_k}_1\in\cR(\rho_1^{1/2})\}\enskip
{\bf\Huge B}\equiv\{\mbox{all state
vectors}\enskip
\ket{f_n}_2\in\cR(\rho_2^{1/2})\}
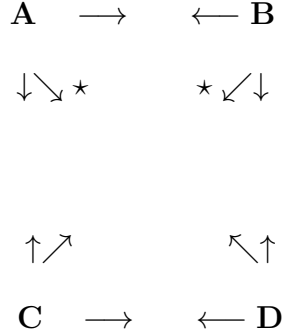
$$

$$\enskip {\bf \Huge
A}\quad\longrightarrow\quad
\quad\longleftarrow {\bf \Huge B}\enskip$$
$$\downarrow\searrow\star\enskip
\quad\qquad\star \swarrow\downarrow$$
\vspace{0.5cm}
$$\enskip\uparrow\nearrow\quad\qquad\qquad
\nwarrow\uparrow$$ $$\enskip {\bf \Huge
C}\quad\longrightarrow\quad
\quad\longleftarrow {\bf \Huge D}$$

$${\bf\Huge C}\equiv \{\mbox{all state
vectors}\ket{\chi_n}_1\in\cR(\rho_1)\}$$
$${\bf\Huge D}\equiv\{\mbox{all state
vectors}\enskip\ket{\phi_k}_2\in \cR(
\rho_2)\}$$

\begin{quote}
\scriptsize \indent {\it Caption.} Each
arrow goes from one of the sets (A,B,C,D)
towards another. It stands for the
corresponding bijection specified in Theorem
2. Oppositely oriented arrows denote
mutually inverse bijections. The diagram is
{\it commutative}, i. e., the successive
bijections combine into the displayed
corresponding one. For instance, taking the
bijection {\bf B}$\downarrow${\bf D} after
the bijection {\bf A}$\rightarrow${\bf B}
gives the bijection {\bf A}$\searrow${\bf
D}, etc.

The imaginary vertical line cutting the
square into two equally wide halves makes
these completely symmetric (due to the
symmetry between the two subsystems in
\$\ket{\Phi}_{12}\$ ).

$\star$ The downward diagonal bijections
{\bf A}$\searrow${\bf D} and {\bf
C}$\swarrow${\bf B} have the {\it physical
meaning} of {\it remote
state
preparations}.\\
\end{quote}

Theorem 2 completes previous work on remote
preparation (or "steering", to use
Schr\"{o}dinger's term) begun by
Schr\"{o}dinger,$^1$ and continued in Ref.
15. To make the completion more precise, the
following theorem clarifies the issue.\\

{\bf Theorem 3}: {\bf A)} A state vector
\$\ket{\phi}_2\$ is obtainable by remote
preparation in \$\ket{\Phi}_{12}\$ if and
only if \$\ket{\phi}_2\in\cR(\rho_2^{1/2})$.

{\bf B)} {\it The set of all} atomic events
\$\ket{j}_1\bra{j}_1\$ the occurrence of
which in nearby-subsystem measurement in
\$\ket{\Phi}_{12}\$  remotely prepares a
given state vector \$\ket{\phi}_2\$ is (in
terms of state vectors \$\ket{f}_1\$ and
\$\ket{g}_1\$):
$$\Big\{\ket{j}_1\bra{j}_1:\ket{j}_1=\alpha
\ket{f}_1+\beta\ket{g}_1\Big\},\eqno{(49a)}$$
where $$\ket{f}_1\equiv U_a^{-1}\rho_2
^{-1/2}\ket{\phi}_2,\eqno{(49b)}$$
$$|\alpha|>0,\quad |\alpha|^2+|\beta|^2=1,
\eqno{(49c)}$$ and $$\ket{g}_1
=Q_1^{\perp}\ket{g}_1\eqno{(49d)}$$ (\$Q_1\$
being the range projector of \$\rho_1,\$ and
\$Q_1^{\perp}\$ being its orthocomplementary
projector, i. e., the null projector),
otherwise \$\ket{g}_1\$ is arbitrary.

{\bf C)} If the occurrence of the atomic
event \$\ket{j}_1\bra{j}_1\$ remotely
prepares \$\ket{\phi}_2,\$ then the {\it
probability} of occurrence is {\it
proportional} to \$|\alpha|^2\$ (cf (49a)).
It is {\it maximal} if and only if
\$\ket{j}_1=\ket{f}_1\$ (cf (49b)).

{\bf D)} If \$\ket{\phi}_2\in\cR(\rho_2)\$
(cf (32)), then
\$\ket{f}_1\in\cR(\rho_1^{1/2}),\$ where
\$\ket{f}_1\$ is given by (49b). Thus, one
has \li remote preparation in this case,
where the maximal probability is the {\it
characteristic weight} of \$\ket{\phi}_2\$
(cf Definition 3).

{\bf E)} If
\$\ket{\phi}_2\in\Big(\cR(\rho_2^{1/2})
-\cR(\rho_2)\Big),\$ where "$-$" denotes
set-theoretical subtraction (of a subset),
then \$\ket{f}_1\in\Big(\bar\cR(\rho_1)-
\cR(\rho_1^{1/2})\Big),\$ where
\$\ket{f}_1\$ is given by (49b).\\

If the ranges of \$\rho_s,\enskip s=1,2\$
are finite dimensional, then the largest
probability is always the characteristic
weight corresponding to \li remote
preparation (cf (32)).

{\it Proof} of Theorem 3: A) The most
general case of remote pure-state
preparation in a bipartite state vector
\$\ket{\Phi}_{12}\$ is given by (13d).
Replacing \$A_a\$ by its polar-factorized
form \$\rho_2^{1/2}U_aQ_1\$ (cf (9b)), one
can see that it is necessary that
\$\ket{\phi}_2\in \cR(\rho_2^{1/2}).\$ That
this is also sufficient is obvious from the
fact that \$\ket{\phi}_2\$ is obtained by
remote preparation when the atomic event
\$\ket{f}_1\bra{f}_1\$ occurs, where
\$\ket{f}_1\$ is given by (49b). (We again
utilize the above polar-factorized form (9b)
of \$A_a\$ in (13d).)

B) Since \$\ket{\phi}_2=p^{-1/2}U_a\rho_1
\ket{j}_1=p^{-1/2}U_a\rho_1Q_1\ket{j}_1\$
(cf (13d) and (9a)), where \$p\$ is the
probability of occurrence, it is obvious
that the occurrence of each of the atomic
events \$\ket{j}_1 \bra{j}_1\$  (cf (49a-d))
remotely prepares \$\ket{\phi}_2.\$ On the
other hand, (49a) with \$\ket{f}_1\in\bar
\cR(\rho_1)\$ is the general form of a state
vector from \$\cH_1,\$ and \$A_a=U_a\rho_1
^{1/2}\$ (cf (9a)) is non-singular on
\$\bar\cR(\rho_1)\$, hence, it follows from
(13d) that if \$\ket{j}_1\$ is not in the
set (49a), then the occurrence of
\$\ket{j}_1\bra{j}_1\$ remotely prepares a
state vector \$\ket{\phi'}_2\$ that is
distinct from \$\ket{\phi}_2$.

C) Substituting in (13c) \$A_aQ_1\$ instead
of \$A_a\$ and \$\ket{j}_1\$ by its form in
(49a), one obtains $$p=|\alpha|^2||\rho_1
^{1/2}\ket{f}_1||^2,\eqno{(50)}$$ which is
independent of \$\ket{g}_2\$ (cf (49a)).
Both claims in Theorem 3C are obvious from
(50).

D) and E) The claims of Theorem 3D  and 3E
follow from the following set-theoretical
insight. One has
$$\bar\cR(\rho_1)=\cR(\rho_1^{1/2})+
\Big(\bar\cR(\rho_1)-\cR(\rho_1^{1/2})\Big),
\eqno{(51)}$$ and $$\cR(\rho_1^{1/2})=
\cR(\rho_1)+\Big(\cR(\rho_1^{1/2})-
\cR(\rho_1)\Big),\eqno{(52)}$$ where \$"+"\$
denote the set-theoretical union of disjoint
sets. The operator \$\rho_1^{1/2}\$ maps the
first term on the rhs of (51) {\it into} the
first term on the rhs of (52). This is seen
from the fact that if \$\ket{j}_1\in
\cR(\rho_1^{1/2}),\$ then \$\exists:\enskip
\ket{k}_1,\enskip
\bra{k}_1\ket{k}_1>0,\enskip \rho_1^{1/2}
\ket{k}_1=\ket{j}_1.\$ Then \$\rho_1^{1/2}
\ket{j}_1=\rho_1\ket{k}_1\in\cR(\rho_1).\$

Actually, \$\rho_1^{1/2}\$ maps
\$\cR(\rho_1^{1/2})\$ {\it onto}
\$\cR(\rho_1).\$ Namely, if \$0\not=
\ket{k}_1 \in\cR(\rho_1),\$ then
\$\exists:\enskip 0\not= \ket{j}_1\$ such
that \$\ket{k}_1=\rho_1\ket{j}_1=\rho_1
^{1/2}\Big(\rho_1^{1/2}\ket{j}_1\Big).\$

Finally, since \$\rho_1^{1/2}\$ maps the lhs
of (51) {\it onto} the lhs of (52) {\it in a
non-singular way}, one easily concludes that
this operator maps the second term on the
rhs of (51) onto the second term on the rhs
of
(52).\hfill $\Box$\\

{\bf \large \noindent VI. CONCLUDING
REMARKS}

There is a very basic and elementary general
claim: Every statement valid for all
bipartite state vectors \$\ket{\Phi}_{12}\in
(\cH_1\otimes\cH_2)\$ is either symmetric in
the two subsystems, or if not, then also the
statement symmetrical to it is always valid.
This comes from the essential symmetry
between \$\cH_1\$ and \$\cH_2\$ (in spite of
the fact that one has to use the two factor
spaces in an ordered way).

The results of this article confirm the
claim that at the very core of entanglement
in any \$\ket{\Phi}_{12}\$ is the {\it
correlated subsystem picture} (see section
2). It consists of statements that appear in
symmetrical pairs: the two reducees
\$\tilde\rho_1\$ and \$\tilde \rho_2\$ are
symmetric (cf (10)) and so are the reducees
of twin observables \$\tilde A_1\$ and
\$\tilde A_2\$ (if one takes algebraic twin
observables, i. e., ones with equal relevant
spactra). The symmetry is in terms of the
{\it antiunitary correlation operator}
\$U_a\$ inherent in \$\ket{\Phi}_{12},\$
which establishes a sort of duality (like
between kets and bras) between the closed
ranges \$\bar\cR(\rho_1)\$ and
\$\bar\cR(\rho_2)$.

The correlation operator connects the
orthogonal decompositions (or spectral
forms) \$\tilde\rho_1=
\sum_ir_i\widetilde{\ket{i}_1\bra{i}_1},\$
\$\tilde
\rho_2=\sum_ir_i\widetilde{\ket{i}_2
\bra{i}_2},\$ to which correspond the
spectral forms of (physical) twin
observables \$\tilde A_1=\sum_ia_i
\widetilde{\ket{i}_1\bra{i}_1}\$ and
\$\tilde A_2=
\sum_ia'_i\widetilde{\ket{i}_2\bra{i}_2}$.
(One should remember that the tilde denotes
that the corresponding operator is reduced
to the range of \$\rho_s,\enskip s=1,2.\$)

In the wider view, when also extended twin
observables are taken into account, or,
equivalently, when one considers generalized
twin observables, which has been elaborated
in this article, one treats the wider class
of linearly independent subsystem state
decompositions \$\rho_1=\sum_n
q_n\ket{\chi_n}_1\bra{\chi_n}_1\$ and
\$\rho_2=\sum_kp_k\ket{\phi_k}_2
\bra{\phi_k}_2\$ along with the relevant
generalized twin observables \$A_1\$ and
\$A_2\$ the measurement of either of which
gives rise to the mentioned state
decomposition on the {\it opposite}
subsystem (remote \li \cd of state), but
{\it not} to the decomposition on the same
subsystem (except in the special case of
proper twin observables). The full
mathematical details and beauty of the
generalized physical meaning of the
correlated subsystem picture is expressed
via the commutative diagrams.

If, following Schr\"{o}dinger,$^1$ one tries
to understand entanglement solely in terms
of disentanglement, i. e., in terms of
remote state decomposition, then one wonders
what is left out from this article.

Restricting ourselves first to the (more
important) complete state decompositions and
pure-state preparation, the following is
omitted.

If the bipartite state vector has infinite
entanglement, i. e., if the dimension of the
two ranges of the respective reduced density
operators is infinite, then even among the
observables for which the basic commutation
\$[A_s,Q_s]=0,\enskip s=1,2,\$ is valid
(range-compatible observables), those for
which the eigenbases of the reducees
\$\tilde A_s, \enskip s=1,2\$ are not
entirely within \$\cR(\rho_s^{1/2})\enskip
s=1,2\$ are left out from remote \li
complete state decomposition. Further,
equally for finite and for infinite
entanglement, if at least one of the reduced
density operators is singular, then the
corresponding commutation
\$[A_s,Q_s]=0,\enskip s=1,2\$ can be
violated by some \$A_s,\$ and the remote
state decompositions caused by the
measurement of such violating observables
are also outside our treatment except in
Theorem 3, which addresses the general case.

Considering only the non-selective aspect of
remote influence, from the physical point of
view it may not be clear why should one
attach more importance to \li remote
complete state decomposition than to the
rest mentioned above. The answer lies in the
selective aspect, when one considers remote
\li pure-state preparation. Theorem 3C makes
it clear that these are the nearby-subsystem
occurrences that in measurement in
\$\ket{\Phi}_{12}\$ have {\it the highest
probability}. This fact singles them out in
importance.

In Theorem 2 and Diagram 2 we have treated
\li remote pure-state preparation as part of
\li remote complete state decomposition.
This is methodologically quite correct. But
in view of the mentioned result in Theorem
3C, physically it is more satisfactory to
reverse the roles of the non-selective and
the selective aspects, and to consider the
former as composed out of the latter. In
other words, perhaps it is physically more
correct to consider remote \li complete
state decomposition as consisting of remote
\li pure-state preparations. Then the
physical importance of the latter is shared
by the former.

One should point out that we have not
considered incomplete remote \li state
decomposition or remote \li mixed-or-pure
state preparation. This is much used in
practice as a step towards complete state
decomposition (towards pure-state
preparation).

We may repeat the remark from the
Introduction that in theoretical physics
mathematics and physics are inextricably
connected and the optimal form of the
former, as a rule, gives physical insight,
often in terms of new physical concepts. The
correlated subsystem picture, by itself a
mathematical concept, which has been further
applied to disentanglement in this article,
leads to insight into the structure of
pure-state entanglement in terms of
generalized (proper and extended) twin
observables. In particular, \li remote
pure-state preparation  appears as the
maximal-probability way of such a remote
effect.

Finally, the largest-probability requirement
in remote pure-state preparation leads to
the conclusion (cf Theorem 3C) that, from
the physical point of view, in case of
infinite-dimensional ranges of
\$\rho_s,\enskip s=1,2,\$ one should
generalize "relevant" (for \li influence)
observables by the weaker requirement of
only range-compatible and
detectably-complete ones (cf Definition 5).

\pagebreak

{\bf \large \noindent Appendix A:}

{\it Proof} that
$$\Big\{\ket{\phi_k}\notin
\mbox{span}\{\ket{\phi_1},\dots
,\ket{\phi_ {(k-1)}},\ket{\phi_{(k+1)}},
\dots , \ket{\phi_d}\},$$ $$k=1,2,\dots
\,d;d\in\mbox{\bf N}\Big\}\quad
\Leftrightarrow$$ $$\Big\{\forall
k:\enskip\ket{\phi_k}\notin \mbox{span}\{
\ket{\phi_1},\dots
,\ket{\phi_{(k-1)}}\}\Big\},\eqno{(A1.1)}$$
where {\bf N} is the set of all natural
numbers.

The first requirement on the set
\$\{\ket{\phi_i}:i=1,\dots ,d\}\#
obviously implies the second one. To prove
the inverse implication, we assume {\it ab
contrario} that the first requirement is
not valid, but the second is. Then there
exists \$k\in${\bf N}, \$1\leq k\leq d\#
such that
$$\ket{\phi_k}=\sum_{i=1}^{(k-1)}\alpha_i
\ket{\phi_i}+\sum_{j=(k+1)}^d\alpha_j
\ket{\phi_j}, \eqno{(A1.2)}$$ all
\$\alpha_i\# and all \$\alpha_j\# complex
numbers. On account of the assumed
validity of the second requirement in
(A1.1), not all \$\alpha_j\# can be zero.
We define \$\bar j\equiv
max\{j:\alpha_j\not= 0\}.\# Then (A1.2)
implies $$\ket{\phi_{\bar j}}=\alpha_{\bar
j}^{-1} \Big(\ket{\phi_k}
-\sum_{i=1}^{(k-1)}\alpha_i\ket{\phi_i}
-\sum_{j=(k+1)} ^{(\bar
j-1)}\alpha_j\ket{\phi_j} \Big)$$ in
contradiction to the assumed validity if
the second requirement.\hfill $\Box$\\

{\bf \large \noindent Appendix B:}

{\it Proof} (of Corollary 1) that every
finite subset \$\{\ket{\phi_1}, \dots
,\ket{\phi_k}\}\# of a \li set (finite or
infinite) is a \li basis in the
$k$-dimensional subspace \$\cS_k\# that it
spans. First we prove the claimed
dimensionality of the span.

{\it Total induction.} We assume that the
dimensionality claim is true for
\$(n-1)<k:\$ \enspace
span$\{\ket{\phi_1},\dots ,
\ket{\phi_{(n-1)}}\}=\cS_{(n-1)}.\# Let
\$\ket{\phi_n}\# be \li of the mentioned
preceding state vectors. Let \$P\# project
onto \$\cS_{(n-1)}.\# One has
$$\ket{\phi_n}=
P\ket{\phi_n}+P^{\perp}\ket{\phi_n},
\eqno{(A2.1)}$$ where \$P^{\perp}\equiv
(1-P),\# and \$P^{\perp}\ket{\phi_n}\#
cannot be zero (cf Definition 1). We
define \$\ket{f_n}\equiv
cP^{\perp}\ket{\phi_n},\# where \$c\# is a
normalization constant, and $$\cS_n\equiv
\mbox{span}\{\cS_{(n-1)},\ket{f_n}\}
\eqno{(A2.2)}$$ is a subspace of \$n\$
dimensions. Since \$\ket{f_n}=c
\Big(\ket{\phi_n}-P\ket{\phi_n}\Big),\#
\$\cS_n\subset\mbox{span}\{\ket{\phi_1},
\dots ,\ket{\phi_n}\}.\# It is obvious
from (A2.1) that \$\ket{\phi_n}\in\cS_n.\#
Hence, \$\mbox{span}\{\ket{\phi_1}, \dots
,\ket{\phi_n}\}\subset\cS_n,\# and,
finally, \$S_n=\mbox{span}\{\ket{\phi_1},
\dots ,\ket{\phi_n}\}$.

Since the claim that the span is a
subspace of that many dimension as the
number of \li state vectors is true for
\$n=1,\# total induction implies that it
is true for any \$n\leq k$.

The uniqueness of the expansion follows
from Corollary 3 is one takes an arbitrary
probability distribution
\$\{p_i:i=1,2,\dots ,k;p_i>0;\sum_{i=1}^k
p_i=1\}\$ and one defines \$\rho\equiv
\sum_{i=1}^kp_i\ket{\phi_i}\bra{\phi_i}$.\\

{\bf \large \noindent Appendix C:} {\it
Proof} of Corollary 3. We show that assuming
(27), each element \$\ket{\chi}\$ from the
range \$\cR(\rho^{1/2})\$ can be expanded in
the \li sequence
\$\{\ket{\phi_i}:i=1,2,\dots
,d\leq\infty\}$.

Let $$\tilde\rho^{-1/2}\ket{\chi}=
\sum_i\beta_i \ket{e_i}.\eqno{(A3.1)}$$
Applying the continuous operator
\$\rho^{1/2},\$ one obtains (cf Lemma B)):
$$\ket{\chi}=\sum_i\beta_i\rho^{1/2}\ket{e_i}=
\sum_i\beta_ip_i^{1/2}\ket{\phi_i},$$
$$\ket{\chi}=\sum_i\alpha_i \ket{\phi_i},
\eqno{(A3.2)}$$ where \$\forall i:\enskip
\alpha_i=\beta_ip_i^{1/2}.\$ On account of
(A3.1) and (28), one has $$\forall
i:\enskip \alpha_i=
p_i^{1/2}\Big[\Big(\bra{e_i}
\tilde\rho^{-1/2}\Big)\ket{\chi}\Big].$$
Substituting \$\bra{e_i}\$ from (30), the
last relation enables us to rewrite (A3.2)
as follows: $$\ket{\chi}=
\sum_ip_i\Big[\Big(\bra{\phi_i}\tilde
\rho^{-1}\Big)\ket{\chi}\Big]
\ket{\phi_i}. \eqno{(A3.3)}$$

Finally, the uniqueness of the expansion
(A3.3) is easily proved by bringing the
opposite assumption into contradiction with
the definition of linear independence (cf
Definition 1).\hfill $\Box$\\

\setlength{\parindent}{1ex}$^1$E.
Schr\"{o}dinger, Proc. Cambridge Phil. Soc.
{\bf 31}, 555 (1935).

\setlength{\parindent}{1ex}$^2$F. Herbut and
M. Vuji\v{c}i\'{c}, Ann. Phys. (NY) {\bf
96}, 382 (1976).

\setlength{\parindent}{1ex}$^3$F. Herbut and
M. Vuji\v{c}i\'{c}, in {\it Proceedings of
the International School

\setlength{\parindent}{2ex} of Physics
"Enrico Fermi"}, Course IL, ed. B.
d'Espagnat (Academic

\setlength{\parindent}{2ex} Press, London,
1971), pp. 316-329.

\setlength{\parindent}{1ex}$^4$G.
Cassinelli, E. De Vito, and A. Levrero, J.
Math. Analys. and Appl. {\bf

\setlength{\parindent}{2ex} 210}, 472
(1997).

\setlength{\parindent}{1ex}$^5$I. M.
Gel'fand and N. Ya. Vilenkin, {\it
Generalized Functions}, Vol. 4

\setlength{\parindent}{2ex} (Academic Press,
New York, 1964).

\setlength{\parindent}{1ex}$^6$J. M. Jauch,
{\it Foundations of Quantum Mechanics}, pp.
175-178 (Addison-

\setlength{\parindent}{2ex} Wesley, Reading,
Mass.,1968).

\setlength{\parindent}{1ex}$^7$F. Herbut and
M. Vuji\v{c}i\'{c}, J. Math. Phys. {\bf 8},
1345 (1967).

\setlength{\parindent}{1ex}$^8$M.
Vuji\v{c}i\'{c} and F. Herbut, J. Math.
Phys. {\bf 25}, 2253 (1984).

\setlength{\parindent}{1ex}$^9$F. Herbut,
e-print quant-ph/0403101.

\noindent$^{10}$F. Herbut, Physical Review A
{\bf 66}, 052321-1-6 (2002).

\noindent$^{11}$N. Hadjisavvas, Lett. Math.
Phys. {\bf 5}, 327 (1981).

\noindent$^{12}$I. C. Gohberg and M. G.
Krein, {\it Introduction to the Theory of
Linear

\setlength{\parindent}{2ex} Nonselfadjoint
Operators}, Translations of Mathematical
Monographs, Vol.

\setlength{\parindent}{2ex} 18 (AMS,
Providence, R. I., 1969).

\noindent$^{13}$P. Busch, P. J. Lahti, and
P. Mittelstaedt, {\it The Quantum Theory of

\setlength{\parindent}{2ex} Measurement},
Lecture Notes in Physics {\bf M2}
(Springer-Verlag, Berlin,

\setlength{\parindent}{2ex} 1991).

\noindent$^{14}$L. P. Hughston, R. Jozsa,
and W. K. Wootters, Phys. Lett. A {\bf 183},

\setlength{\parindent}{2ex} 14 (1993).

\noindent$^{15}$F. Herbut and M.
Vuji\v{c}i\'{c}, J. Phys. A {\bf 20}, 5555
(1987).

\end{document}